\documentclass[conference]{IEEEtran}
\usepackage{color}
\usepackage[english]{babel}
\usepackage{graphicx}
\usepackage{epstopdf}
\usepackage{times}
\usepackage{multirow}
\usepackage[algo2e,linesnumbered]{algorithm2e} 
\usepackage{algorithmic}
\usepackage{algorithm}
\usepackage{psfrag}
\usepackage[shell]{gnuplottex}
\usepackage{mathenv}
\usepackage{array}
\usepackage[T1]{fontenc}
\usepackage{array}
\usepackage{mdwmath}
\usepackage{multicol}
\usepackage{dblfloatfix}
\usepackage{amsmath}
\usepackage{amssymb}
\usepackage{float}

\usepackage{url}
\makeatletter
\newcommand*{\rom}[1]{\expandafter\@slowromancap\romannumeral #1@}
\makeatother

\begin{document}

\title{Dynamic Channel Allocation for Interference Mitigation in Relay-assisted Wireless Body Networks}

  \author{\IEEEauthorblockN{Mohamad Jaafar Ali\IEEEauthorrefmark{1}, Hassine Moungla\IEEEauthorrefmark{1}, Ahmed Mehaoua\IEEEauthorrefmark{1}, Yong XU\IEEEauthorrefmark{2}}
\IEEEauthorblockA{\IEEEauthorrefmark{1}LIPADE, University of Paris Descartes, Sorbonne Paris Cit\'{e}, Paris, France \\
}{\IEEEauthorrefmark{2}School of Computer Science and Engineering, South China University of Technology, China}, \\Email: \{mohamad.ali; hassine.moungla; ahmed.mehaoua\}@parisdescartes.fr}

\maketitle
\begin{abstract}
%\boldmath
We focus on interference mitigation and energy conservation within a single wireless body area network (WBAN). We adopt two-hop communication scheme supported by the the IEEE 802.15.6 standard 2012 \cite{key26}. In this paper, we propose a dynamic channel allocation scheme, namely DCAIM to mitigate node-level interference amongst the coexisting regions of a WBAN. At the time, the sensors are in the radius communication of a relay, they form a relay region (RG) coordinated by that relay using time division multiple access (TDMA). In the proposed scheme, each RG creates a table consisting of interfering sensors which it broadcasts to its neighboring sensors. This broadcast allows each pair of RGs to create an interference set (IS). Thus, the members of IS are assigned orthogonal sub-channels whereas other sonsors that do not belong to IS can transmit using the same time slots. Experimental results show that our proposal mitigates node-level interference and improves node and WBAN energy savings. These results are then compared to the results of other schemes. As a result, our scheme outperforms in all cases. Node-level signal to interference and noise ratio (SINR) improved by 11dB whilst, the energy consumption decreased significantly. We further present a probabilistic method and analytically show the outage probability can be effectively reduced to the minimal.

\end{abstract}
\IEEEpeerreviewmaketitle

\section{Introduction}
A WBAN is a wireless network of wearable sensors that may be embedded inside or attached on the human body. The sensors are used in various applications such as health monitoring, ubiquitous healthcare, sports and entertainment. These networks mainly monitor vital signs as glucose percentage in blood, heart beats, respiration, body temperature and/or can record electrocardiography (ECG) \cite{key2,key14,key21}.

WBANs are more likely to coexist with each other due to their dynamic and mobile nature. Interferences can happen due to the absence of coordination amongst them. Also, concurrent transmissions of a particular WBAN can interfere with each others as well as introduce interferences at some other nodes (intra-WBAN interference). Firstly, radio interference is of the paramount importance since it can quickly degrade a WBAN's performance. Secondly, the stringent factor is energy and so this requires to always keep as low power consumption as possible. However, the latest version of the IEEE 802.15.6 standard proposed in February 2012 for WBANs supports two-hop communication and states that up to 10 co-located WBANs can function properly within a transmission range of 3 meters \cite{key20, key21}.

Due to the constrained nature of WBANs in terms of energy, size and cost, advanced antenna techniques cannot be used for interference mitigation as well as power control mechanisms used in cellular networks are not applicable to WBANs. Thus, novel methods and schemes are required for interference mitigation and consequently for energy conservation in both intra- and inter-WBANs. 

The rest of the paper is organized as follows. Section \rom{2} presents the recent works addressed problems related to interference mitigation and energy conservation for WBANs. Section \rom{3} describes the network model and presents a virtual relaying scheme for WBANs. Section \rom{4} shows the proposed dynamic channel allocation for interference mitigation (DCAIM) scheme and also presents a propabilistic approach to reduce outage probability. Section \rom{5} shows and explains the experimental results. The conclusions and future works are drawn in section \rom{6}. 

\section{Related Work}
Recent studies show multi-hop scheme has a lower power consumption in comparison to one-hop scheme. Using relays reduces the whole WBAN interference and its power consumption \cite{key7,key8,key10}. On the other hand, other works prove that TDMA scheme is an attractive solution to avoid interference within an intra-WBAN. These schemes require time synchronization which is infeasible when large number of WBANs coexist \cite{key1,key3,key21}. Some efforts focus on co-channel interference analysis of non-overlapping WBANs. They develop a model for efficient network planning and resource management by using geometrical probability approach \cite{key4}. Other research works assume one desired and multiple coexisting interferers WBANs. The coordinator of the desired WBAN calculates SINR periodically and afterwards commands its nodes to select an appropriate interferece mitigation scheme \cite{key1}. 

Other studies as in \cite{key6} analyze performance parameters of a reference WBAN in the presence of another interfering network. Moreover, these parameters have been improved by adoption of an optimized time hopping code assignment strategy. Works in \cite{key5} consider a WBAN where the coordinator periodically polls its sensors. The nodes however calculate SINR then select the modulation scheme according to the experienced channel quality. Authors of \cite{key7} propose a single-relay cooperative scheme. Where a set of relays computes individually the required transmission power to participate in the cooperative communication. Eventually, the best relay is selected in a distributed fashion. 

Furthermore, authors of \cite{key6} explore the problem of interference among multiple coexisting WBANs. The proposed scheme enables two or three WBANs to agree on a common TDMA schedule. This scheme reduces the whole interference amongst them. The work in \cite{key3} investigates the problem of coexisting WBANs. It adopts a TDMA scheme within a WBAN and a carrier sensing mechanism to deal with inter-WBAN interference. The coordinator checks the channel whether free before polling the sensor to avoid inter-WBANs interference.

Authors of \cite{key8} consider a sensor network with slot based transmission scheme. This work provides an analytical framework to optimize the size of relay-zone around each source node. Authors of \cite{key16} propose a resource allocation scheme for interference mitigation among multiple coexisting WBANs. Each pair of interfering WBANs form an interference region. However, the nodes belonging to this region are later allocated orthogonal sub-channels. Whilst, other nodes that do not belong to this region transmit using the same time slot.

Most of the recent works show problems related to interference mitigation and power conservation in inter-WBANs environment. They do not address problems related to \textbf{node-level} interference and energy consumption within an intra-WBAN. However, in this work, we are interested to address these problems. Our proposed DCAIM scheme exploits the benefits of multi-hop communication schemes for interference mitigation and energy conservation within a single WBAN. Thus, it allows the high interfering nodes of a single WBAN to transmit through using orthogonal channels whilst, others nodes also with acceptable low levels of interferences to transmit using the same channel.

\section{Network Model}
\subsection{Virtual Network Model}
\begin{figure}[ht]
  \centering
        \includegraphics[width=0.4\textwidth]{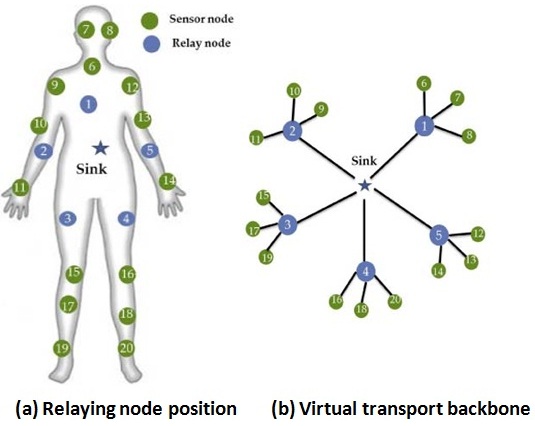}
\caption{A WBAN-based relaying virtual backbone}
\label{fig:virtualbackbone}
\end{figure}

To address these interference and energy concerns, we propose in this work a reliable topology design for WBANs as shown in Fig. \ref{fig:virtualbackbone}. that takes into account the mobility of the patient while guaranteeing a reliable data delivery. We consider a tree-based multi-hop WBAN topology where in addition to sink and sensor nodes, we introduce a set of relay nodes distributed along the body to serve as transport network for sensory and control data. We call this relaying structure virtual backbone network where each node has some fixed neighbors to which it has constant Euclidian distance independently of body gestures. Accordingly, this backbone structure has relatively stable and reliable links in the presence of postural mobility related to body movement.
\subsection{Network Assumptions and Description}
More specifically, we consider a single WBAN formed of a set of sources $N_{s}$, set of relays $N_r$ and coordinator C. We adopt two-hop communication (source - relay - C) as an energy efficient scheme. The WBAN is divided into multiple relay regions (RGs) as shown in Fig. \ref{fig:wban}. Where each region is formed of a set of sources $S_{s}$ and a set of relays $R_{r}$. Any region is coordinated by a relay. This requires that the sources are in the radius communication of a relay. Whilst, the relay transmits sources' data to a common destination C. Thus, in our model we assume the followings:
\begin{itemize}
  \item The underlying sources of each RG use TDMA to communicate with the relay
  \item Any TDMA schedule of any specific RG may overlap with other RGs' schedules
  \item The relays compete for the shared channel using slotted CSMA/CA to communicate with the sink
  \item Within each RG, all the sources are in the radius communication of at least one relay
  \item Within each RG, the sources are distributed based on the application requirements in use
  %\item Sources transmit large size data packets
 \end{itemize}
\subsection{Channel Model}
The radio wave propagation in a WBAN is noticeably different compared to the other environments since it takes place inside or close to the human body. WBAN channel modeling is challenging due to the complex shape and composition of human body. In WBANs, the small-scale fading is dominant compared to the large-scale fading due to the short distance between the transmitter and the receiver. It is claimed that the lognormal distribution is the most fitting to describe the fading channel of WBANs since the human body is the dominating shadowing factor. Furthermore, the lognormal distribution is the best choice for modeling the small-scale fading in WBANs because there are only a small number of multi-path components from the diffraction around the body \cite{key21}.

The path loss (PL) can be described in the Eq. \ref{eq: 1} below:
\begin{equation}\label{eq: 1}
PL(d) = PL(d_{0}) + 10 \times \log_{10} \left( \frac{d}{d_{0}}\right)^{\alpha} + X_{\sigma} 
\end{equation}
Where
\begin{itemize}
\item PL(d): is the path loss at distance d from the transmitter
\item $PL(d_{0})$: is the path loss at reference distance $d_{0}$ from the transmitter
\item $\alpha$: is the path loss exponent
\item $ X_{\sigma}$: is a log normal distributed random variable
\end{itemize}
\IncMargin{1em}
\begin{algorithm}
\footnotesize
\SetKwData{Left}{left}\SetKwData{This}{this}\SetKwData{Up}{up}
\SetKwFunction{Union}{Union}\SetKwFunction{FindCompress}{FindCompress}
\SetKwInOut{Input}{input}\SetKwInOut{Output}{output}
%\Input{A}
%\Output{B}

Step 1: Orthogonal Transmission 

    \For{$i\leftarrow 1$  $\KwTo$  $N_{r}$}
       { 
        \For{$j\leftarrow 1$  $\KwTo$  $S_{s}$}
          {
            Sensor (i, j) is transmitting
            
            $RG_{k}$ estimates the received signal power, $\delta_{i, j}$ 
            
            from sensor (i, j)
          }
           
        }
      
Step 2: Determine the inter-interference set
      
      \For{$i\leftarrow 1$  $\KwTo$  $N_{r}$}
       { 
          $\delta_{min, i}$ = min\{$\delta_{i, j}$\}
          
          \For{$k\leftarrow 1$  $\KwTo$  $N_{r}$, k $\neq$ i}
            { 
             \For{$j2\leftarrow 1$  $\KwTo$  $S_{s}$}
               {
                  \If{$\delta_{k, j2}$ > $\delta_{min, i}$ - $\delta_{Thr}$}
                   {
                      Add (k, j2) to set $IL_{i}$
                
                   }
               
                }
                
            }
            
        }
       
Step 3: Broadcast
  \For{$i\leftarrow 1$  $\KwTo$  $N_{r}$}
     {
  
      Broadcast $IL_{i}$ 
  
     }
     
Step 4: Determine the interference set
  
    \For{$i\leftarrow 1$  $\KwTo$  $N_{r}$}
     {

      $IS_{i}$ = $IL_{i}$ $\cup$ \{(i,j) | (i,j) $\in$ $IL_{k}$, k $\neq$ i\}

     }

Step 5: Channel assignment

  \For{$i\leftarrow 1$  $\KwTo$  $N_{r}$}
      {
      
      Leave the time slots for nodes in $IS_{i}$ unchanged
      
       Equally assign the remaining channels to nodes that do
       
       not belong to $IS_{i}$ 
        
      }

\caption{Dynamic Channel Assignment for Interference Mitigation (DCAIM)}
\label{dcaim}
\end{algorithm}
\DecMargin{1em}
 
\subsection{Dynamic Channel Assignment Proposed Scheme}
Inter-RGs interference can happen due to the concurrent transmissions of sources of different RGs. Based on our assumptions, a source transmission of a specific RG may introduce higher interference level on the nodes of other nearby RGs whilst other sensors of that RG may have a very low interference on the surrounding RGs. A simple approach to avoid inter-RGs interference is to assign orthogonal channel to those nodes which experience higher interference levels.

We consider a total of $N_r$ RGs are co-located within a WBAN, each RG consists of $S_{s}$ source nodes. The shared channel among the sources of any RG is divided evenly into $S_{s}$ time-slots. On the first side, we assume that $k^{th}$ source sensor of $RG_{i}$ transmits to its relay using its time slot $T_{i,k}$. On the other side, other relays try to compute the interference level of that source based on the power received from that signal. 
     
\subsubsection{Interference Lists}
Let $\delta_{i, j, k}$ denote the received power from $k^{th}$ source of $RG_{j}$ at $RG_{i}$. After the orthogonal transmissions of sources of all RGs are complete in the first round. This enables each relay to compute and maintain a table consisting of the received powers from each source of all RGs. Also, each relay finds the minimum received power from sensors within its $RG_{i}$ which denoted by $\delta_{min, i}$ = min\{$\delta_{i, j, k}$\}. It then compares it to the received power from sources of other RGs. We present the interference list of $RG_{i}$ as $IL_{i}$, where $IL_{i}$ = \{(j,k)|$\delta_{i, j, k}$ > $\delta_{min, i}$ - $\delta_{Thr}$\}. The relay $R_{i}$ of $RG_{i}$ includes a sensor from other RGs in its $IL_{i}$ if its received power is larger than $\delta_{min, i}$ - $\delta_{Thr}$.
\begin{figure}[ht]
  \centering
        \includegraphics[width=0.4\textwidth]{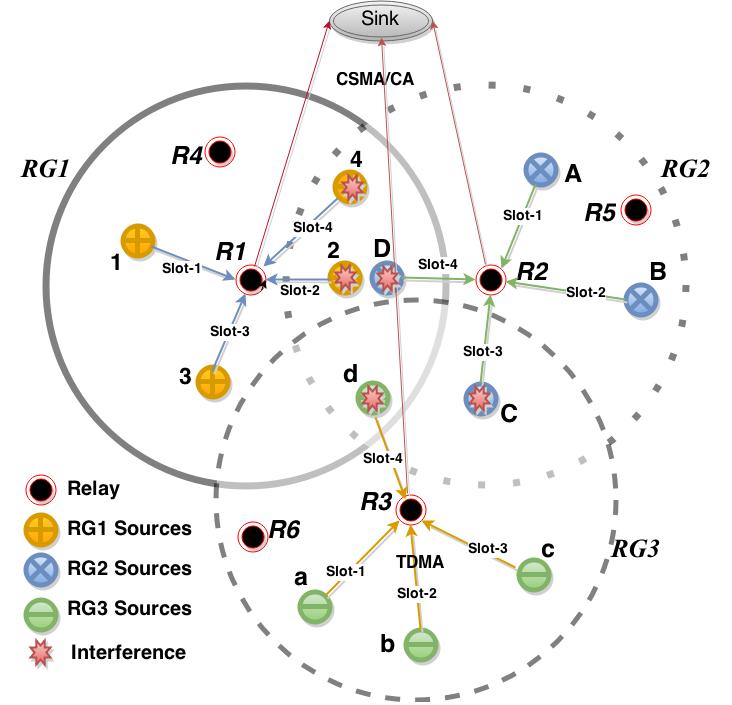}
\caption{An example of distribution of nodes amongst three RGs and their
Interference Region}
\label{fig:wban}
\end{figure}

\subsubsection{Interference Sets}
Based on interference lists broadcasts amongst relays, each relay can verify which of its sources introduce severe interference on other RGs. And also, which sources of other RGs impose severe interference on its RG. Afterwards, each relay creates an interference set formed of all sources that have a severe interference level. We denote interference set of $RG_{i}$ by  $IS_{i}$ = $IL_{i}$ $\cup$ \{(i,k) | (i,k) $\in$ $IL_{j}$, j $\neq$ i\}
\begin{figure}[ht]
  \centering
        \includegraphics[width=0.4\textwidth]{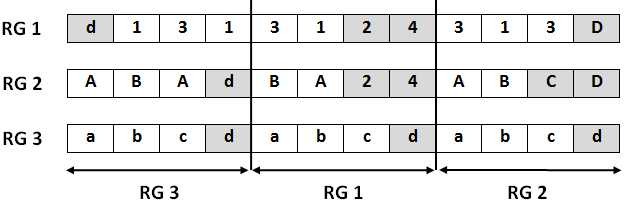}
\caption{Proposed channel assignment for nodes in Fig. \ref{fig:wban}}
  \label{fig:slots}
\end{figure}
\subsection{Channel Assignment Mechanism - Illustrative Example}
Three coexisting relay regions RG1, RG2 and RG3 of a WBAN are shown in Fig. \ref{fig:wban}. We present an illustrative example to clarify the channel assignment mechanism in our proposed Algo. \ref{dcaim} for typical scenario of a WBAN. Each circumference of a circle denotes the coverage range of each RG and the interference region amongst them is represented by the intersection of their coverage ranges. The interference list of each $RG_{i}$ is as follows:
\begin{itemize}
\item $IL_{1}$ = \{(2,D), (3,d)\}
\item $IL_{2}$ = \{(1,2), (1,4), (3,d)\}
\item $IL_{3}$ = \{(2,C)\}
\end{itemize}
Generally, $IL_{ID1}$ =  \{(ID2,ID3), ... \} where ID2 is the identifier of RG interfering with the specified $RG_{ID1}$ and ID3 is the identifier of the node of the $RG_{ID2}$ which is interfering with RG of interest $RG_{ID1}$. Each relay broadcasts its Interference list (IL) and based on other RGs braodcasts, it creates its interference set (IS). Doing so, each RG forms an interference set as follows:
\begin{itemize}
\item $IS_{1}$ = \{(2,D), (3,d), (1,2)\}
\item $IS_{2}$ = \{(1,2), (1,4), (3,d), (2,C), (2,D)\} 
\item  $IS_{3}$ = \{(2,C),  (3,d)\}
\end{itemize}

However, each relay of RG assigns orthogonal channels for each of its nodes that are in the interference set of other co-located RGs. Whilst, other nodes are allowed to use all the free time slots until they are all occupied. Therefore, each node will be allocated and assigned more slots which increases the overall spatial reuse. As an example, Fig. \ref{fig:slots} shows that node 1 of $RG_{1}$ have been assigned four slots. That means four times more spatial reuse than the case where each node is assigned only one slot. It is clear to see in Fig. \ref{fig:slots} that the nodes of interfering sets (ISs) are not transmitting concurrently. This ensures the fact that the interference level is kept as low as possible. 
The following pseudo code shown in Algo. \ref{dcaim} clarifies the different steps of the the DCAIM proposed scheme.
\begin{table}
\centering
\normalsize
 \begin{tabular}{ |p{3.5cm}|p{3.5cm}|}
\hline
\multicolumn{2}{|c|}{Parameters List} \\
\hline
transmission power & -10 dBm \\\hline
sensitivity & -84.7 dBm   \\\hline
noise floor & -102 dBm \\\hline
interference distance d   & 10 cm  \\\hline
data rate & 250 kbps  \\\hline
base frequency & 2.4 GHz  \\\hline
path loss exponent &  4.22 \\\hline
sub-channels &  8 channels \\ \hline
\end{tabular}
 \caption {Simulation parameters} \label{tab:parmlist}
\end{table}
\section{DCAIM Proposed Scheme Analysis - Outage Probability}
We denote the probability that the total interference at time instant i is larger than $\delta_{Thr}$ at $RG_{0}$ by $P_{out}^{i}$. Then, we calculate this probability by the following formula:
\begin{equation}
P_{out}^{i} = \left(\displaystyle\sum_{j=1}^{N-1} \delta_{j} > \delta_{Thr} \right)
\end{equation}
We present a probabilistic approach which we prove analytically it lowers the outage probability. As mentiond above, any sensor whose received SINR at $RG_{0}$ is higher than a threshold is added to the interference region of $RG_{0}$. We denote by $\delta_{j}$ the received SINR from a sensor in $RG_{j}$ at $RG_{0}$. An orthogonal channel is assigned to that sensor with certain probability which equals $\frac{\delta_{i}}{\delta_{Thr}}$. Thus, at time instant i, we can calculate the average interference level using the proposed probabilistic approach as follows:
\begin{equation}
\delta_{i} = \displaystyle\sum_{j=1}^{N-1} \delta_{j} \left(1 - \frac{\delta_{j}}{\delta_{Thr}} \right)
\end{equation}
Based on the probabilistic approach, any sensor with probability $\frac{\delta_{i}}{\delta_{Thr}}$ is assigned with an orthogonal channel.

\textbf{Lemma 1:} We denote by $P_{Probilistic}$ and $P_{Original}$ the outage probability of probabilistic approach and the outage probability of the original scheme respectively. Then, $P_{Probilistic}$ < $P_{Original}$. If we let the average reuse factor $R_{prob}$ and $R_{0}$ for the proposed scheme with and without probabilistic channel assignment respectively, then $R_{prob}$ <  $R_{0}$.

\textbf{Proof:} Based on outage probability definition, we have:
\begin{equation} 
P_{Probilistic} = p \left(\displaystyle\sum_{i=1}^{N-1} \delta_{i} \left(1 - \frac{\delta_{i}}{\delta_{Thr}} \right) > \delta_{Thr}\right) 
\end{equation}
\begin{equation} 
= p \left(\displaystyle\sum_{i=1}^{N-1} \delta_{i} > \delta_{Thr} + \displaystyle\sum_{i=1}^{N-1} \frac{\delta_{i}^2}{\delta_{Thr}}\right) 
\end{equation}  
\begin{equation}             
< p  \left( \displaystyle\sum_{j=1}^{N-1} \delta_{j} >  \delta_{Thr} \right) = P_{Original}  
\end{equation}
Where the last line of $P_{Probilistic}$ is based on the fact that the CDF is an increasing function of its argument. We define $P_{probabilistic, I, i}$ as the probabilistic approach deployment probability that a sensor node of $RG_{i}$ is in the interference region of $RG_{0}$. Then:
\begin{equation} 
P_{prob, I, i} = P(\delta_{i} > \delta_{Thr}) + P(\delta_{i} < \delta_{Thr})\frac{\delta_{i}}{\delta_{Thr}},   
\end{equation}
which is greater than $P_{i}$ = $P(\delta_{i} > \delta_{Thr})$. To complete the proof, let $P_{I, i}$ as the probability that a sensor node of $RG_{i}$ exists in the interference region of $RG_{0}$. Then:
\begin{equation} 
 R_{0} = N - \left( \displaystyle\sum_{i=1}^{N-1} P_{I, i}\right)    
\end{equation} 
To complete the proof, the average reuse factor of probabilistic approach is calculated as the following: 
\begin{equation} 
R_{prob} = N -  \displaystyle\sum_{i=1}^{N-1} P_{prob, I, i}
\end{equation}
\begin{equation} 
= N - \displaystyle\sum_{i=1}^{N-1} P_{I} - \displaystyle\sum_{i=1}^{N-1} P(\delta_{i} < \delta_{Thr})E \left( \frac{\delta_{i}}{\delta_{Thr}} \right) 
\end{equation}
\begin{equation} 
< N - \displaystyle\sum_{i=1}^{N-1} P_{I} = R_{0}   
\end{equation} 
Where $P_{I}$ denotes the probability that the received power is above the threshold value $\delta_{Thr}$
\section{Experimental Results}
In our simulation scenario, we consider a WBAN network consisting of three relay regions ($N_{r} = 3$). Each RG consisting of set $R_{r} = 2$ relay nodes and set $S_{s} = 4$ source sensors. The WBAN is modeled for on-body communication and is located within an area of 1x2 $m^{2}$. We set the same transmission power for all the nodes in the network except the coordinator. For simplicity, the simulation parameters are listed in Table \ref{tab:parmlist} above.

\subsection{WBAN Energy Consumption}
\begin{figure}[ht]
  \centering
        \includegraphics[width=0.4\textwidth]{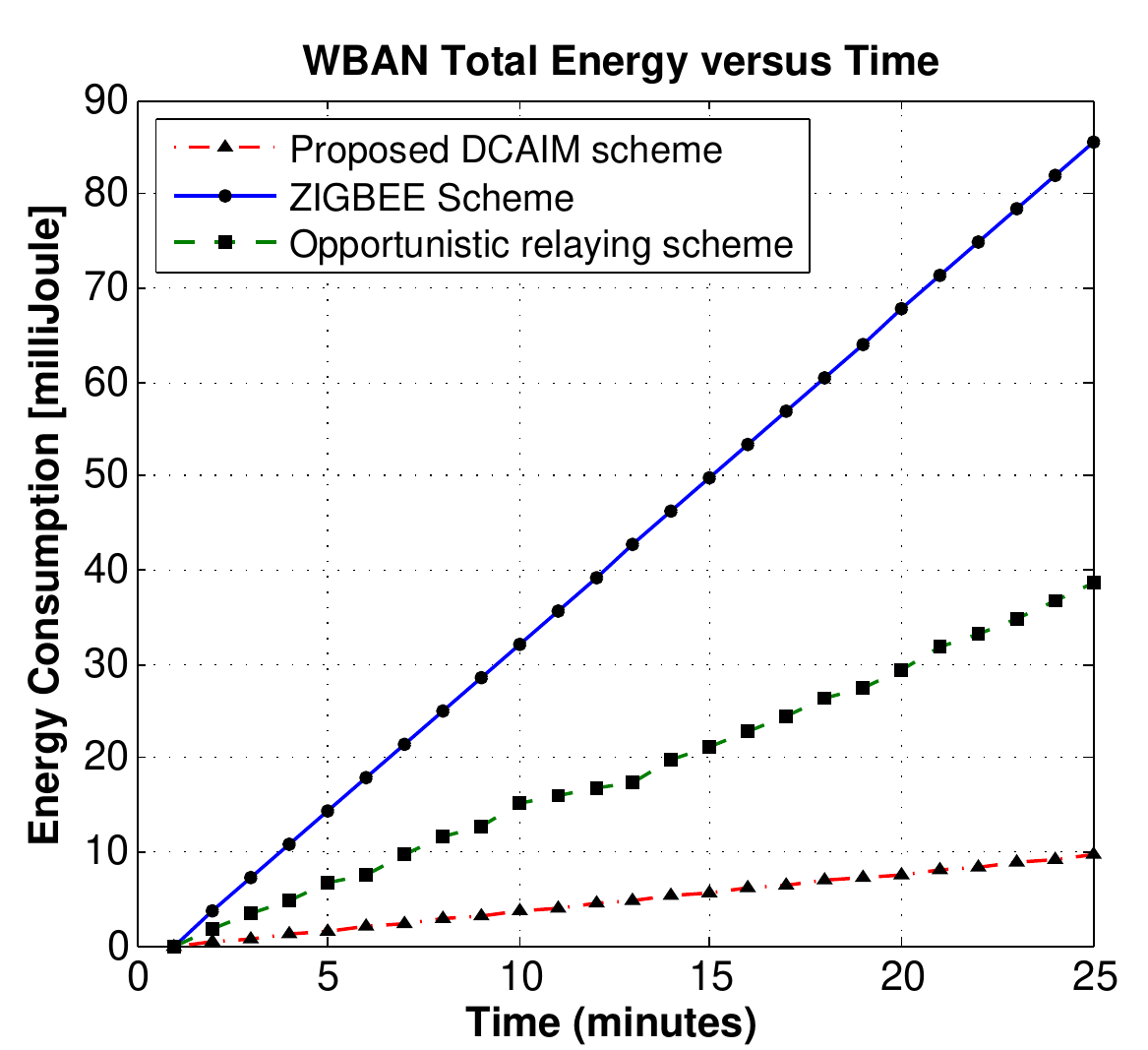}
\caption{Sum of energy consumption of all WBANs versus time for all different schemes}
  \label{fig:dcaim}
\end{figure}

As shown in Fig. \ref{fig:dcaim}, we compare the sum of energy consumption of each WBAN versus time for all WBANs of the same simulation setup. The first WBAN implements our proposed DCAIM scheme, the second implements CSMA/CA access mechanism with opportunistic relaying (OR) scheme and the third WBAN allows its sources to transmit their data directly to common coordinator using the traditional Zigbee MAC scheme (one-hop). However, as can be clearly seen in this figure, firstly, the energy consumption of DCAIM scheme is always less than the energy consumption of the other schemes. Secondly, a lower energy consumption is obtained for both two-hop schemes than single-hop scheme. It can be evidently seen from the same figure that the energy conumption of our proposed DCAIM scheme increases slightly versus time, whilst in the other both schemes increases sharply, specifically, in Zigbee MAC scheme. This is because each high interfering node is assigned orthogonal channel to avoid interference. Hence, the collisions and retransmissions are reduced. Doing so makes less interference introduced on other nodes and consequently conserves their energy. However, this justifies that our DCAIM proposed scheme outperforms other schemes and extends longer a WBAN's energy lifeime. 
\subsection{Nodes Signal to Interference and Noise Ratio (SINR)}
\begin{figure}[ht]
  \centering
        \includegraphics[width=0.4\textwidth]{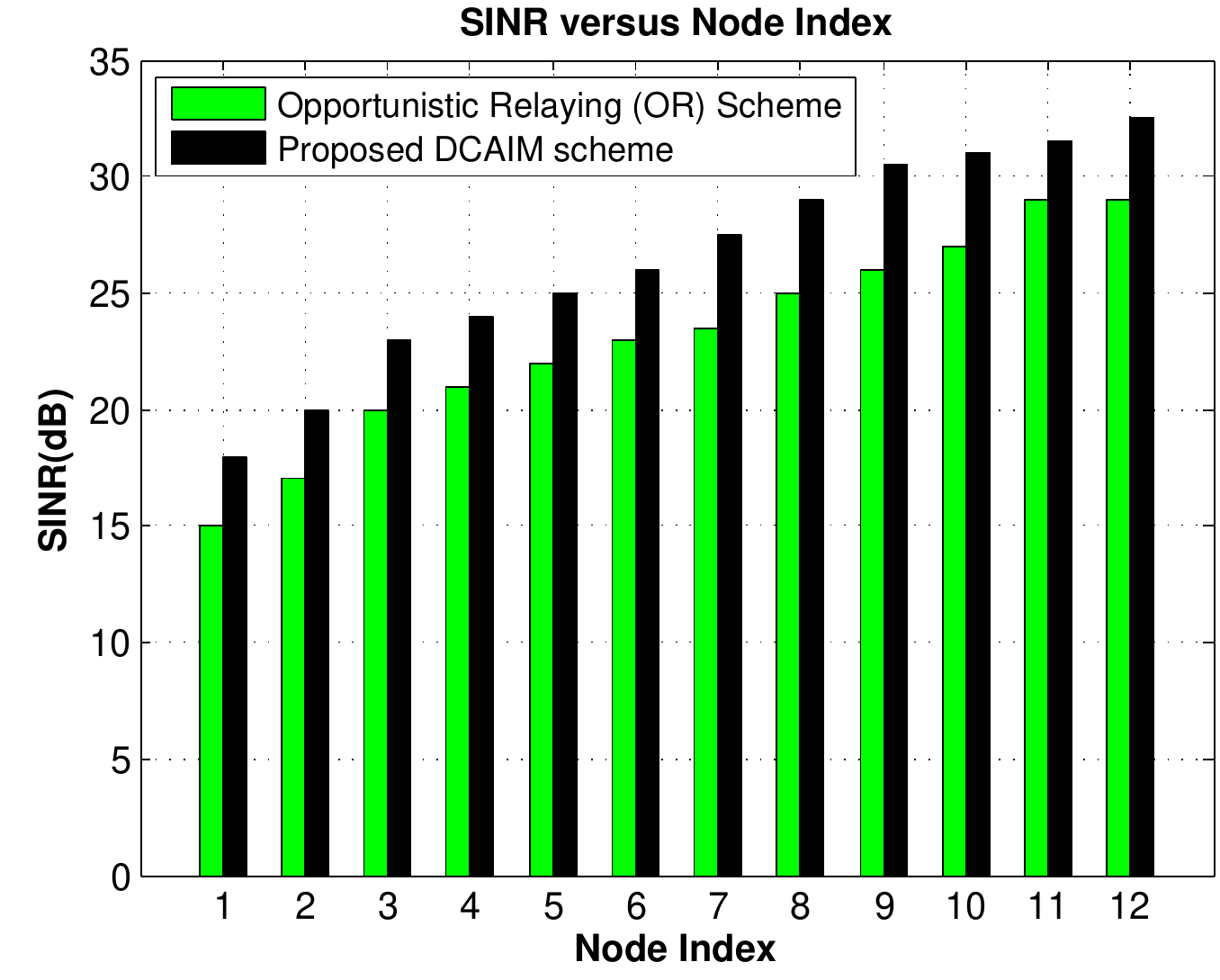}
\caption{SINR versus node index of each WBAN for DCAIM and OR schemes}
  \label{fig:sinr}
\end{figure}

As shown in Fig. \ref{fig:sinr} and based on the same simulation setup, we compare the node's SINR results of a WBAN that implements our proposed DCAIM scheme with another that implements OR scheme. Both schemes adopt two-hop communication. However, it can be clearly seen from this figure that SINR results of our proposed scheme outperforms and always larger than the SINR values of OR scheme at each node. In other words, the interference is mitigated at each node level by improving its SINR. These nodes which were supposed to introduce interference on others nodes, after our proposed scheme, they are assigned orthogonal sub-channels. Doing so, lets them avoiding interference and consequently increases their SINR. Also, our proposed scheme increases the SINR by 11dB on average. This justifies that our proposed scheme mitigates node-level interference and hence this decreases the energy consumption of each node and WBAN as whole as well as extends its lifetime. 
\section{Conclusion}
A distributed and dynamic channel allocation scheme for interference mitigation (DCAIM) is presented in this paper to mitigate node-level interference amongst coexisting regions of a WBAN. The proposed scheme states that the sensor nodes that introduce significant interferences on other RGs are assigned orthogonal channels whereas other nodes can use the same time slots allocated before. The simulation results show that our proposed DCAIM  scheme can reduce the interference at both node and WBAN levels and hence extend WBAN lifetime. We further compare the simulation results of our proposal with other scheme using OR and Zigbee single-hop scheme, the results show that ours outperforms in terms of interference mitigation and power savings in all cases. Additionally, we study and evaluate our proposed approach by theoretical analysis and show the outage probability can be effectively reduced to the minimal which allows to better and efficient use of the limited resources of WBANs. However, the study of our proposed scheme for a dynamic WBAN topolopy is left as future research issue.

%\bibliography{mybib}{}
%\bibliographystyle{plain}

\end{document}